\documentclass[12pt]{article}

\usepackage{epsfig}
\usepackage{amsmath,amssymb,amsfonts}
\usepackage{float,a4wide}
\newcommand{\be}{\begin{equation}}
\newcommand{\en}{\end{equation}}
\newcommand{\bea}{\begin{eqnarray}}
\newcommand{\ena}{\end{eqnarray}}

\newcommand{\hbo}{\hbox to 1 true cm {\hfill } }

\newcommand{\tr}{\text{tr}}

\newcommand{\Tr}{\text{Tr}}

\def\dslash{\partial\kern-.5em\slash}
\def\kslash{k\kern-.5em\slash}
\def\pslash{p\kern-.5em\slash}
\def\Dslash{D\kern-.5em\slash}

\begin{document}
%
%\vglue 1truecm

\vbox{ KA-TP-7-2003 
\hfill UNITU-THEP-05/03
}

\vfil
\centerline{\large \bf Vortex structures in pure SU(3) lattice gauge theory } 

\vspace{0.5cm}

\bigskip
\centerline{\large  Kurt Langfeld } 
\bigskip
\vspace{.5 true cm}
{\small
\centerline{Institut f\"ur Theoretische Physik, Universit\"at
   Karlsruhe  }
\centerline{D-76128 Karlsruhe, Germany}
\bigskip

\centerline{ and } 
\bigskip

\centerline{ Institut f\"ur Theoretische Physik, Universit\"at
   T\"ubingen }
\centerline{D-72076 T\"ubingen, Germany}
}
\vspace{1.5 true cm}
\centerline{ October 10, 2003 }
\vspace{.5 true cm}

\vskip 1.5cm

\begin{abstract}
\noindent
The structures of confining vortices which underlie pure $SU(3)$ 
Yang-Mills theory are studied by means of lattice gauge
theory. Vortices and $Z_3$ monopoles are defined as  dynamical
degrees of freedom of the $Z_3$ gauge theory which emerges by center
gauge fixing and by subsequent center projection.  It is observed for
the first time for the case of $SU(3)$ that these degrees of
freedom are sensible in the continuum limit: the planar vortex density
and the monopole density properly scales with the lattice spacing. 
By contrast to earlier findings concerning the gauge group $SU(2)$, 
the effective vortex theory only reproduces 62\% of the full string
tension. On the other hand, however, the removal of the vortices 
from the lattice configurations yields ensembles with vanishing 
string tension. $SU(3)$ vortex matter which originates from  Laplacian 
center gauge fixing is also discussed. Although these vortices 
recover the full string tension, they lack a direct interpretation 
as physical degrees of freedom in the continuum limit.
\end{abstract}

\vfil
\hrule width 5truecm
\vskip .2truecm
\begin{quote}
\end{quote}
\eject
\section{Introduction }
\label{intro}

Since QCD was recognized as the theory of strong interactions 
by means of high energy scattering experiments, the question arose 
as to whether QCD also explains the absence of quarks in the 
particle spectrum. After extensive numerical simulations had become 
feasible with modern computers, it became clear that the pure gluonic 
theory already bears witness to quark confinement: the static quark 
anti-quark potential rises linearly for large distances due to 
the formation of a color-electric flux tube~\cite{Bali:1994de}. 
Moreover, the low-energy model, in which a fluctuating bosonic string 
plays the role of the effective degree of freedom, predicts a 
characteristic $1/r$ correction to the potential at large distances. 
This picture recently received viable support from lattice simulations 
which verified the dependence of this term on the number of dimensions 
at a quantitative level~\cite{Luscher:2002qv}. 
A major challenge of modern quantum field theory is the question: 
Why does the color-electric flux tube form? 

\vskip 0.3cm
Over the recent past, lattice gauge simulations have strengthened the
idea that topological degrees of freedom, which are characteristic 
for the non-Abelian nature, are relevant for confinement. 
Among those, color-magnetic monopoles and center vortices are under 
intense discussion (for a most recent review
see~\cite{Greensite:2003bk}). Here, we will focus 
on the vortex picture of confinement. 

\vskip 0.3cm
The central idea is to simplify pure Yang-Mills theory under the 
retention of its confining capability, hoping to filter 
out degrees of freedom which meet two criteria: (i) the degrees of
freedom are sensible in the continuum limit of lattice gauge theory, 
and (ii) they are closely related to confinement of pure Yang-Mills 
theory. 

\vskip 0.3cm
Gauge fixing and projection techniques have proven to be convenient 
for these purposes. Center gauges have been designed 
to maximize the importance of center degrees of freedom, and 
the projection $SU(N) \longrightarrow Z_N$ was proposed for the 
simplification process~\cite{DelDebbio:1996mh,DelDebbio:1998uu}. 
Vortices and $Z_N$ monopoles appear as the dynamical degrees of
freedom of the $Z_N$ gauge theory. Indeed, criteria (i) and (ii) were
seen to be satisfied in the case 
of a $SU(2)$ gauge theory if the maximal center gauge (MCG) is 
adopted~\cite{DelDebbio:1996mh,DelDebbio:1998uu,Langfeld:1997jx,Langfeld:2001nz}. 

\vskip 0.3cm
Of particular interest is the case of the $SU(3)$ gauge group because 
of its relevance to the theory of strong interactions.
In the present paper, a thorough study of the confining vortices 
is performed for the gauge group $SU(3)$. For the first time the
planar vortex density as well as the density of the $Z_3$ monopoles
are reported to properly extrapolate to the continuum limit in the
case of MCG.  In sharp contrast to the case of $SU(2)$, the vortices
fail to recover the string tension to its full extent. 
We will see below, however, that there is still a close 
relation of the MCG vortices to confinement: removing the 
vortices yields a model theory with vanishing string tension.
These findings will be contrasted to those obtained 
in the Laplacian center gauge.

\vskip 0.3cm
The techniques for extracting static quark potentials as well as 
the numerical setup are explained in the next section. 
Details of the definition and construction of the vortex matter 
are given in section \ref{text}. To which extent the 
vortex matter is able to reproduce the $SU(3)$ string tension 
is studied in section \ref{dom}. Thereby, new high-precision data 
for the case of $SU(2)$ are presented. These data serve as a 
``contrast agent'' for the findings for the $SU(3)$ case. 
In section \ref{properties}, the properties of vortices 
and $Z_3$ monopoles are discussed in the continuum limit. 
Conclusions are left to the final section.

\section{ The static quark potential }
\label{pot}

Focal points are the study of the relevance of the vortices for the
$SU(3)$ string tension and of the properties of the vortex matter
emerging in the continuum limit. For these purposes, a careful 
determination of the physical value of the lattice spacing is
mandatory. This is the subject of the present section.

\subsection{Numerical Setting }
\label{setting}

Results of the simulation of pure $SU(N)$ gauge lattice gauge theory
will be presented below for $N=2$ and $N=3$, respectively. 
The dynamical degrees of freedom are the unitary matrices 
$U_\mu (x) \in SU(N)$. Configurations $\{U_\mu (x)\}$ will 
be generated according to the Wilson action 
\bea
S &=& \beta \; \sum _{x, \mu < \nu } \; \frac{1}{2N} \, \tr \Bigl( 
P_{\mu \nu }(x) + P^\dagger _{\mu \nu }(x) \Bigr) \; , 
\\ 
P_{\mu \nu }(x) &=& U_\mu (x) \; U_\nu (x+\mu) \; 
U^\dagger_\mu (x+\nu) \; U^\dagger_\nu (x) \; , 
\label{pla}
\ena 
where $P_{\mu \nu }(x)$ is the plaquette. The update is 
performed using the Creutz heat-bath algorithm~\cite{Creutz:zw}
for the case of $SU(2)$. Updating the diagonal $SU(2)$ 
subgroups as proposed by Cabibbo and Marinari~\cite{Cabibbo:zn} 
is performed in the case of $SU(3)$. 
Each ten heat-bath sweep is accompanied 
by four micro-canonical reflections in order to reduce 
autocorrelations. Measurements were taken after twenty such
blocks of sweeps. Most of the data are taken on a $L^4$, $L=16$
lattice. In the case of the $SU(2)$ gauge group, lattices with 
$L=24$ were also studied. 

\vskip 0.3cm 
The static quark anti-quark potential $V(r)$ will be extracted 
from planar, rectangular Wilson loops $W(R,T)$ of extension $
R \times T$, i.e., 
\be
\Bigl\langle W(R,T) \Bigr\rangle \; \propto \; 
\exp \bigl\{ - V(r) \; a \; T \bigr\} \; , \hbo 
r \; := \; R\, a , \hbo T : \; \; \mathrm{large} \; , 
\label{wil} 
\en 
where $a$ is the lattice spacing.

\subsection{ Overlap enhancement } 
\label{enhance}

In order to extract the physical signal from noisy Wilson loops, 
so-called overlap enhancement has been proven to be 
an important tool~\cite{Albanese:ds,Teper:wt,Bali:ab}. 
For the 
case of $SU(2)$\footnote{ For the case of $SU(3)$, a novel method 
will be proposed below.}, we closely 
follow the procedure in~\cite{Bali:ab} and define the cooled 
{\it spatial} links  by  
\be 
\Pi _i (x) \; = \; {\cal P}_N \; \sum _k \; U_k(x) \; U_i (x+k) \; 
U^\dagger _k (x+i) \; , \hbo i=1,\,  2,\, 3\, , 
\label{smear} 
\en 
where $k$ runs from $-3 \ldots 3$ and $k=i$ is excluded from the sum. 
$ {\cal P}_N$ is the projector onto the 'closest' 
$SU(N)$ element. In the case of $SU(2)$ and for 
$M = a_0 + i \vec{\tau} \vec{a}$, $\tau ^a$ being the Pauli matrices, 
the effect of the operator  ${\cal P}_2$ is 
\be 
{\cal P}_2 M \; = \; \frac{1}{\sqrt{a_0^2 + \vec{a}^2}} \; 
\Bigl( a_0 + i \vec{\tau} \vec{a} \Bigr) \; . 
\en 
\begin{figure}
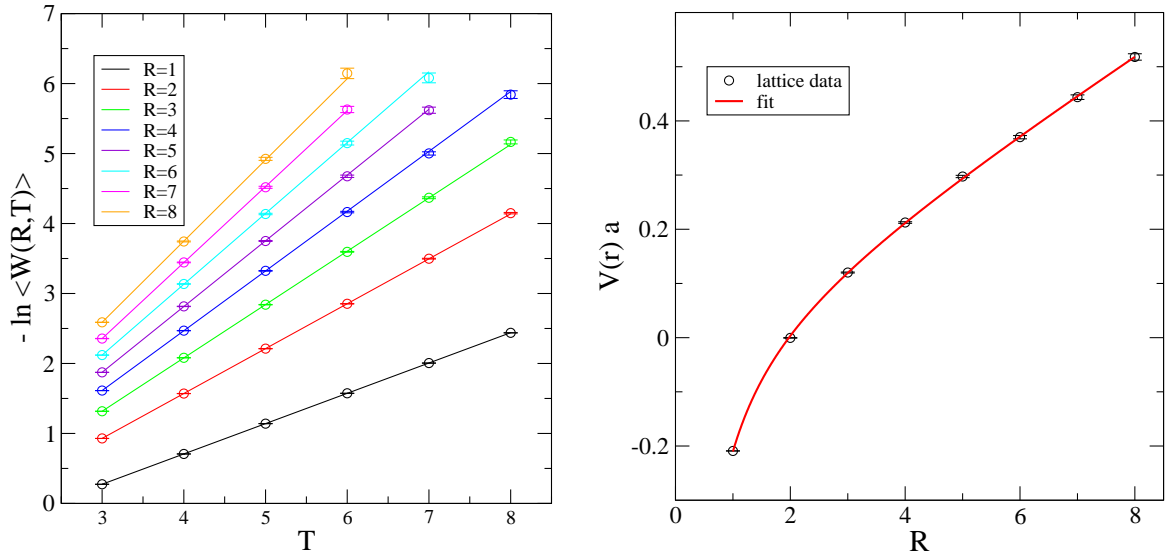

\begin{center} 
\epsfig{figure=wrt.eps,width=7.7cm}
\epsfig{figure=pot59.eps,width=8cm}
\end{center}
\caption{Linearity of the function $-\ln \; \langle W(R,T) \rangle $ 
in $T$ for several values of $R$ (left panel). The static 
quark anti-quark potential for $SU(3)$, $16^4$ lattice and 
$\beta = 5.9$ (right panel). }
\label{fig:1}
\end{figure} 
Details concerning ${\cal P}_3$ can be found in~\cite{Bali:ab}. 
Temporal links are unchanged, i.e., 
\be 
\Pi _0 (x) \; = \; U_0(x) \; . 
\en 
Since the projection ${\cal P}_N$, $N=3$ is ``expensive'' from a numerical 
point of view, we used a different method to define the 
cooled spatial links $\Pi _i(x)$, $i=1,2,3$. Let us define 
the action of the spatial links of a given time slice $t$ by 
\be 
S_{(3)} (t) \; = \; \sum _{i>k, 1 \ldots 3} P_{ik}(x) \; , 
\label{act3}
\en 
where $P_{ik}(x)$ (\ref{pla}) is the plaquette calculated from 
the spatial links. In addition, we define three different embeddings 
of the $SU(2)$ matrix $a_0 +  i \vec{\tau} \vec{a}$, $a_0^2 +
\vec{a}^2 = 1$  into the group $SU(3)$, i.e., 
\be
V^{(1)} \; = \;  \left( \begin{matrix}
 1 & 0 & 0 \\ 
0 & a_0 + a_3 & a_2 -i a_1 \\
0 & -a_2 -i a_1 & a_0 - a_3 
\end{matrix}  \right) \; , \;  \;  \;  \;  
V^{(2)} \; = \; \left( \begin{matrix}
a_0 + a_3 & 0 & a_2 -i a_1 \\
 0 & 1 & 0 \\ 
-a_2 -i a_1 & 0 & a_0 - a_3 
\end{matrix}  \right) \; , 
\en 
\be 
V^{(3)} \; = \; \left( \begin{matrix}
a_0 + a_3 & a_2 -i a_1 & 0 \\
 -a_2 -i a_1 & a_0 - a_3 & 0 \\ 
0 & 0 & 1 
\end{matrix}  \right) \; . 
\en 
Let us now consider a particular spatial link $U_l(x)$. 
Substituting $U^\prime := V^{(1)} U_l(x)$, we locally maximize the action 
$S_{(3)}$ with respect to  $V^{(1)}$. Subsequently, we replace 
$ U^\prime $ by $ U^{\prime  \prime  } = V^{(2)} U^\prime $ and 
maximize with respect to $ V^{(2)} $, and setting 
$ U^{\prime  \prime \prime  } = V^{(3)} U^{\prime  \prime  }$, 
$ V^{(3)}$ is chosen to maximize $S_{(3)}$. Finally, we define 
\be 
\Pi _l (x) \; = \; V^{(3)} \; V^{(2)}  \; V^{(1)}  \; U_l(x) \; . 
\label{smear2} 
\en 
We then visit the next link on the lattice. One sweep has been
performed when all spatial links of the lattice have been visited. 
The advantage of the present procedure is that the maximization 
of $S_{(3)}$ with respect to one of the $SU(2)$ subgroups can 
be implemented very efficiently. 

\begin{figure}
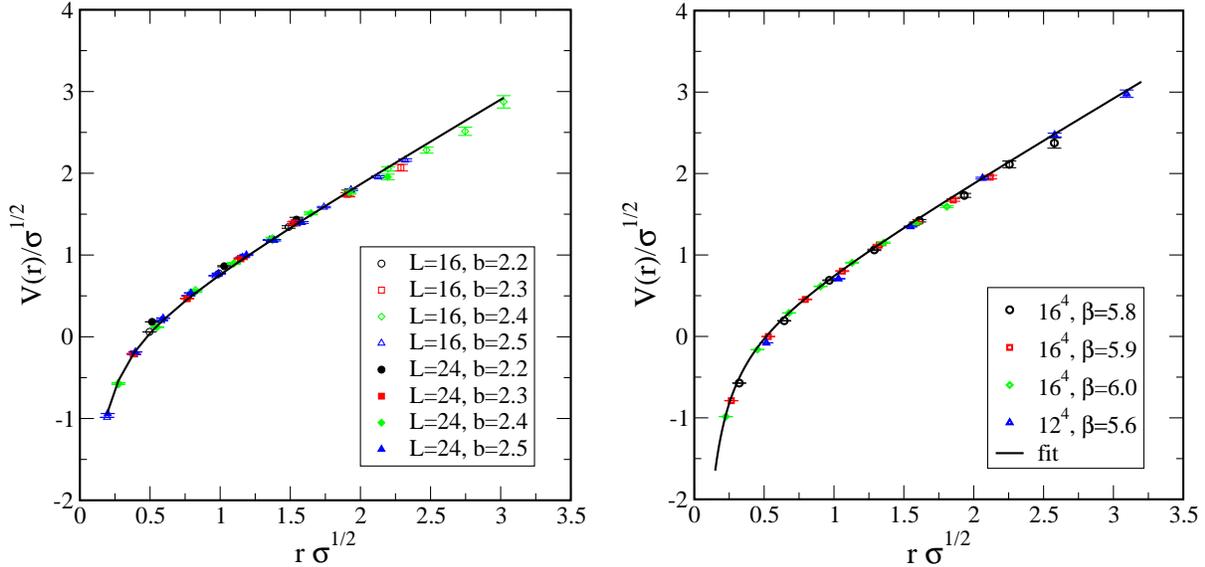

\begin{center} 
\epsfig{figure=pot_full_su2.eps,width=8cm}
\epsfig{figure=pot_full3.eps,width=8cm}
\end{center}
\caption{ The static quark anti-quark potential for the $SU(2)$ 
gauge group (left panel) and for the case of the $SU(3)$ 
gauge group (right panel).  }
\label{fig:2}
\end{figure} 
\vskip 0.3cm 
In order to achieve a good overlap with the groundstate, the above 
procedure for determining the links $\Pi _l(x)$ is applied 
recursively, and the Wilson loop expectation value $\langle W(R,T) 
\rangle $ is calculated from the configurations $\{ \Pi _\mu (x) \}$ 
rather than the ensembles $\{ U _\mu (x) \}$ . 
The average of $S_{(3)}$ over the time slices divided 
by the total number of spatial links serves as litmus paper 
for the overlap. It turned out that ten sweeps are enough to yield 
more than $0.99$ ground state overlap. Good overlap is 
also signaled by the quantity 
\be 
- \ln \; \Bigl\langle W(R,T) \Bigr\rangle \; , 
\label{lnwil}
\en 
which already shows a linear behavior in $T$ for $T \ge 3$. 
This is illustrated for the case of a $SU(3)$ gauge group, 
$L^4=16^4$, and $\beta =5.9$. The final results are obtained 
from 100 independent measurements. The symbols in figure \ref{fig:1}, 
left panel, represent the lattice data of the quantity  (\ref{lnwil});
the lines are linear fits in $T$, i.e., 
\be 
- \ln \; \Bigl\langle W(R,T) \Bigr\rangle \; = \; \gamma T \; + \; \delta 
\; . 
\en 
The coefficients $\gamma $ can be interpreted as $V(r) a$. The latter 
quantity is shown in figure \ref{fig:1}, right panel (symbols). The line 
is a fit according to the function 
\be 
V(r) \; a \; = \; c \; R \; - \; \frac{b}{R} \; + \; V_0 \; , 
\hbo r = R \, a\; , 
\en 
where the parameter $c$ can interpreted as the string tension in 
units of the lattice spacing, i.e., $c= \sigma a^2$. 
This method for the calculation of the string tension was 
used, e.g., in~\cite{Stack:sy}. For the present example, we find 
\be 
\sigma \; a^2 \, ( \beta = 5.9 ) \; = \;  0.0701 \; \pm \;  0.0005 \;
, 
\en 
which is in agreement with the value reported in 
the literature, i.e., $ \sigma \; a^2 \, ( \beta = 5.9 ) 
= 0.073$ ~\cite{Fingberg:1992ju,Bali:1992ru}.

\subsection{ The scaling relation } 
\label{scaling} 

In order to mark a dimensionful quantity as a physically sensible one 
in the continuum limit $a \rightarrow 0$, it is crucial 
to express this quantity in units of a physical reference scale. 
Throughout this paper, the string tension $\sigma $ will serve 
this purpose. 

\vskip 0.3cm 
It is time consuming but straightforward to determine the dependence of 
$\sigma \;a^2(\beta )$ on $\beta $. The results for the gauge group 
$SU(2)$ are summarized in table \ref{tab:1}. Those 
for the case of a $SU(3)$ gauge group are shown in 
table \ref{tab:2}. 

\begin{table}[t]
\begin{center}
{\small 
\begin{tabular}{c|c|c|c|c|c|c|c|c} \hline \hline 
L & 16 & 24 & 16 & 24 & 16 & 24 & 16 & 24\cr \hline 
$\beta $ & 2.2 & 2.2 & 2.3 & 2.3 & 2.4 & 2.4 & 2.5 & 2.5\cr \hline 
$\sigma a^2 $ & 0.26(2) & 0.24(1) & 0.146(3) & 0.145(2) & 0.0752(7) &  
0.0754(5) & 0.0391(3) & 0.0373(1) 
\cr \hline \hline 
\end{tabular} 
}
\end{center}
\caption{ String tension $\sigma $ in units of the lattice spacing $a$
for the case of the $SU(2)$ gauge group. } 
\label{tab:1}
\end{table}

\begin{table}[h]
\begin{center}
{\small 
\begin{tabular}{c|c|c|c|c|c|c} \hline \hline 
L & 12 & 16 & 16 & 16 & 16 & 16 \cr \hline 
$\beta $ & 5.6 & 5.6 & 5.7 & 5.8 & 5.9 & 6.0 \cr \hline 
$\sigma a^2 $ & 0.32(1) & 0.26(1) & 0.169(3) & 0.104(1) & 
0.0701(5) & 0.0514(3) 
\cr \hline \hline 
\end{tabular} 
}
\end{center}
\caption{ String tension $\sigma $ in units of the lattice spacing $a$
for the case of the $SU(3)$ gauge group. } 
\label{tab:2}
\end{table}

These values are in good agreement with those reported, e.g., 
in~\cite{Fingberg:1992ju}.

\goodbreak 

\section{The vortex texture }
\label{text}

Concerning the criteria (i) and (ii) of the introductory section in 
\ref{intro}, 
one firstly notes that the properties of the $SU(2)$ vortices
extrapolate  properly to the continuum limit~\cite{Langfeld:1997jx} 
if the maximal center gauge (MCG) is used. Moreover, the MCG vortex
theory correctly reproduces the de-confinement 
temperature~\cite{Langfeld:1998cz}. The phase transition acquires a 
geometrical picture: is appears as a vortex de-percolation phase 
transition~\cite{Langfeld:1998cz,Engelhardt:1999fd} which already 
points towards 
a weak vortex interaction (see also~\cite{Engelhardt:1999wr}). 
Secondly, it was observed that this geometrical picture correctly 
reproduces the finite size scaling of the 3D Ising universality 
class~\cite{Langfeld:2003zi}. This shows that at least for
temperatures close to de-confinement, MCG vortices interact weakly. 

\vskip 0.3cm
It turns out that the non-locality induced by gauge fixing 
is crucial for property (i) above: it was analytically shown that if 
unfixed lattice configurations are projected onto vortex 
configurations, the complete static quark potential (including the 
Coulomb term) is obtained~\cite{Faber:1998en}. At the same time, 
the properties of this vortex matter strongly depend on the 
size of the lattice spacing. Thus, these vortices lack an 
interpretation in the continuum limit. In these respects the 
non-locality of the approach to the vortex matter is an advantage. 
On the other hand, however, this non-locality generically 
makes it difficult to access the vortex matter in practical 
simulations: using a variational gauge condition as, e.g., in the 
case of MCG~\cite{DelDebbio:1998uu}, the definition of the vortices 
is ambiguous as a result of the inability to localize the 
global maximum of a non-linear functional. At least for small lattice 
sizes, the so-called Gribov ambiguity might have a significant 
influence on physical
observables~\cite{Bornyakov:2000ig,Bertle:2000py}. 

\vskip 0.3cm
Marked progress concerning the Gribov problem 
was made with the construction of the Laplacian
gauges~\cite{Vink:1992ys}: the non-locality of the gauge fixing 
is preserved while the maximization is replaced by an 
eigenvector problem, which can be handled by present-day 
algorithms. The Laplacian version of the MCG was firstly proposed 
for a $SU(2)$ gauge group in~\cite{Alexandrou:1999iy} and generalized 
to the $SU(3)$ case in~\cite{deForcrand:2000pg}. A further
improvement was reported in~\cite{Faber:2001zs}. Vortex 
matter of the Laplacian center gauge (LCG) is unambiguously defined 
and recovers the asymptotic string tension for both gauge groups,
$SU(2)$ and $SU(3)$. In the case of a $SU(2)$ gauge theory, it was,
however, observed that the LCG vortices are produced in large 
abundance, implying that they lie dense (nevertheless in a controlled
way) in the continuum limit~\cite{Langfeld:2001nz}. 
This indicates a rather strong interaction of LCG vortices, which 
might render it difficult to mimic LCG vortex matter in a low-energy 
effective model. 

\vskip 0.3cm
The present section briefly reviews details of the 
center gauge fixing procedures and the vortex projection 
techniques with an emphasis on the case of $SU(3)$.

\subsection{ The ideal center vortex cluster }

In order to reveal degrees of freedom which are relevant for 
confinement, we are looking for configurations $\{Z_\mu (x) \}$, 
$Z_\mu (x) \in Z_N$ which best represent the full link configurations 
$\{U_\mu (x) \}$, $U_\mu (x) \in SU(N)$. Thereby, $Z_N$ represents 
the center of the group $SU(N)$, i.e., 
\be 
Z_\mu (x) \; = \; \exp \Bigl\{ i \frac{ 2 \pi }{N} \; m \Bigr\} \; , 
\hbo - \frac{N}{2} < m \le  \frac{N}{2} \; , 
\label{zelem}
\en 
where $m$ is an integer. 
There is an optimal choice of the gauge for which the overlap 
of the center configurations with the full ones is maximal. 
Let us denote the gauge transformed links by 
\be 
U^\Omega _\mu (x) \; = \; \Omega (x) \; U_\mu (x) \; 
\Omega ^\dagger (x+\mu) \; , \hbo \Omega (x) \in SU(N) \; . 
\label{gauge}
\en 
In oder to obtain the ideal center configurations, we minimize the 
functional 
\be 
\sum _{x, \mu } \Bigl[ U^\Omega _\mu (x) \; - \; Z_\mu (x)
  \Bigr]^\dagger \; \Bigl[ U^\Omega _\mu (x) \; - \; Z_\mu (x)
  \Bigr]  \;  \;   \stackrel{ \Omega, Z_\mu }{\longrightarrow } \;  \;  
\mathrm{min} 
\label{acmin}
\en 
with respect to $\Omega (x) $ and $Z_\mu (x) $. The selection 
of $\Omega (x)$ implies the choice of a gauge. We will call this 
gauge {\it ideal center gauge } (ICG) throughout this paper. 
The condition (\ref{acmin}) directly implies that the 
overlap, i.e., 
\be 
R \; = \; \frac{1}{N_l} \; \sum _{x, \mu } \; \Re  \; \frac{1}{N} \; \tr 
\Bigl(  U^\Omega _\mu (x) \;  Z^\dagger _\mu (x) \Bigr) \; , 
\label{overlap}
\en 
is maximized. $N_l$ is the number of links of the lattice, and 
$-1 \le R \le 1$. $R=1$ means that the link configuration 
$\{U_\mu (x) \}$ can be entirely expressed in terms of center elements 
after a suitable gauge has been chosen.

\vskip 0.3cm 
The maximization of $R$ with respect to the center elements 
$Z_\mu (x)$ can be performed locally: with 
\be 
\frac{1}{N} \; \tr  U^\Omega _\mu (x) \; = \; u_l \, \exp \{ i \varphi ^l
\}, \hbo  Z_\mu (x) \; = \; \exp \{ i \varphi ^l_m \} \; , 
\hbo \varphi _m = \frac{ 2 \pi }{N} \; m  \; , 
\label{with}
\en 
where $l = \{x, \mu \}$ specifies the link,  one finds 
$$
R \; = \; \frac{1}{N_l} \; \sum _{l } \; u_l \; \cos \Bigl( \varphi ^l 
- \varphi ^l _m \Bigr) \; . 
$$
The optimal choice is obtained by choosing $m$ (\ref{with}) in such a 
way that 
$\varphi ^l _m $ is closest to $\varphi ^l $. 
This mapping 
\be 
SU(N) \rightarrow Z_N : \; \;  U^\Omega _\mu (x) \rightarrow 
Z_\mu (x)  \; = \; {\cal Z} \Bigl( U^\Omega _\mu (x) \Bigr) 
\label{project}
\en 
is called center projection. Inserting $Z_\mu (x)  $ from
(\ref{project}) into (\ref{overlap}),  the overlap $R$ must then be 
maximized with respect to the gauge transformations, i.e., $\Omega
(x)$.

\subsection{ Center gauges } 
\label{mcg} 

Since the mapping (\ref{project}) by no means depends smoothly 
on $\Omega (x)$, an iteration over-relaxation algorithm which 
iteratively determines $Z_\mu (x)$ and $\Omega (x)$ from 
(\ref{overlap}) can hardly work. State of the art would be to 
determine the desired quantities $Z_\mu (x)$, $\Omega (x)$ by the 
technique of ``simulated annealing''. However in this case,
determining, 
e.g., $\Omega (x)$ to the precision which is needed for the 
vortex analysis is extremely costly from a numerical point of view. 
This approach is beyond the scope of the present paper and is 
left to future investigations. 

\vskip 0.3cm 
In order to make the gauge fixing efficient by means of 
iteration over-relaxation, we assume that in the optimal case,  
$1/N \; \tr U^\Omega _\mu (x) $ comes close to a center element. In this 
case, we relax the condition which constrains  
${\cal Z} ( U^\Omega _\mu (x) ) $ in (\ref{project}) 
to a center element. There are two possibilities for doing this:  
\bea
{\cal Z} \Bigl( U^\Omega _\mu (x) \Bigr) & \propto & 
\Bigl( \tr U^\Omega _\mu (x) \Bigr)^\dagger \; , 
\label{meson} \\ 
{\cal Z} \Bigl( U^\Omega _\mu (x) \Bigr) & \propto & 
\Bigl( \tr U^\Omega _\mu (x) \Bigl)^2 \; . 
\label{baryon} 
\ena 
Hence, we find the gauge conditions 
\bea 
R_{mes} &=& \frac{1}{N_l} \; \sum _{x, \mu }  \; 
\Bigl[ \frac{1}{N} \; \tr \Bigl(  U^\Omega _\mu (x)  \Bigr) \;  
\Bigr] \; 
\Bigl[ \frac{1}{N} \; \tr \Bigl(  U^\Omega _\mu (x)  \Bigr) \;  
\Bigr]^\dagger  \; , 
\label{mmcg} \\ 
R_{bar} &=& \frac{1}{N_l} \; \sum _{x, \mu } \; \Re  \; 
\Bigl[ \frac{1}{N} \; \tr \Bigl(  U^\Omega _\mu (x)  \Bigr) \;  
\Bigr]^3 \; ; 
\label{bmcg} 
\ena 
both have been advertised in the literature~\cite{Faber:1999sq}. 
It is not clear at the beginning which of the above possibilities 
yields the larger overlap (\ref{overlap}). It might turn out that even 
a background-field-dependent admixture of both possibilities 
(\ref{meson},\ref{baryon}) is best for the present purposes. 
Since the algorithm for the so-called ``mesonic'' gauge 
(\ref{mmcg}) was already studied in the literature to a 
large extent~\cite{Faber:1999sq,Montero:1999by}, 
we will employ the gauge condition $R_{mes} \rightarrow max$ 
for the determination 
of $\Omega (x)$ and  perform subsequent center projection 
along the lines outlined in the previous subsection. 

\vskip 0.3cm 
Even once we have agreed on one of the suboptimal gauge conditions 
(\ref{mmcg},\ref{bmcg}), there is still the problem of the Gribov 
ambiguities: since the overlap $R$ is a non-linear functional on $\Omega
(x)$, detecting the global maximum of $R$ is practically impossible 
for reasonable lattice sizes. The choice of the 
local maximum of $R$ which is implicitly defined by the algorithm 
determines the gauge. Although the physics extracted in the 
latter gauge can be highly relevant for confinement, the definition of 
gauge by the numerical procedure is unsatisfactory. 
The Gribov problem can be solved by adopting the Laplacian 
version~\cite{Vink:1992ys,Alexandrou:1999iy,deForcrand:2000pg,Alexandrou:2001fh}of the maximal center gauges. A detailed study of these gauges 
and the corresponding vortex matter can be found for the case 
of $SU(2)$ in~\cite{Langfeld:2001nz} and for the case of 
$SU(3)$ in~\cite{deForcrand:2000pg}. Here, we will briefly outline 
the procedure for the case of $SU(3)$ and refer the reader to
reference~\cite{deForcrand:2000pg} for details. 

\vskip 0.3cm 
The generators $t^a$ of the $SU(N)$ algebra satisfy the equation 
\be 
t^a _{ki} \; t^a_{lm} \; = \; \frac{1}{2} \Bigl( 
\delta _{il} \; \delta _{mk} \; - \; \frac{1}{N} \delta _{lm} \; 
\delta _{mk} \Bigr) \; , \hbo \tr \; t^a t^b \; = \; \frac{1}{2} \delta
^{ab} \; . 
\label{alge}
\en 
With the help of this identity, the ``mesonic'' functional $R_{mes} $ 
(\ref{mmcg}) can be written as 
\be 
R_{mes} \; = \;  \frac{1}{N_l} \; \sum _{x, \mu }  \; 
\frac{1}{N^2} \; \Bigl[ \tr  \; O^T (x) \; R_\mu (x) \; 
O(x+\mu)  \; + \; 1 \Bigl] \; , 
\label{mcgad}
\en 
where the adjoint matrices  are defined by 
\be 
 O ^{ab} (x) \; = \; 2 \, \tr \Bigl\{ t^a  \; \Omega (x) 
 \; t^b  \; \Omega ^\dagger (x) \Bigr\} \; , \hbo 
 R_\mu ^{ab} (x) \; = \; 2 \, \tr \Bigl\{ t^a  \; U_\mu (x) 
 \; t^b  \;  U_\mu ^\dagger (x) \Bigr\} \; . 
\label{adj}
\en 
Equation (\ref{mcgad}) can be written as 
\be 
N^2 \; R_{mes} \; = \;  \frac{1}{N_l} 
\Tr \; {\cal O}^T \; {\cal R} \;  {\cal O} \; + \; 1 \; , 
\en 
where we have introduced, e.g., the vector $ {\cal O} $ of the 
combined coordinate and color space, $\{O^{ab}(x)\} \rightarrow {\cal
  O}$. Up to a term proportional to the unit matrix, 
${\cal R}$ is the adjoint Laplacian operator, i.e., 
\be 
{\cal R}^{ab}_{xy} \; = \; \frac{1}{2} \; \sum _\mu \bigl[ 
R_\mu ^{ab}(x) \; \delta _{y, x+\mu } \; + \; 
R_\mu ^{ba}(x-\mu) \; \delta _{y, x-\mu } \; \Bigr] \; . 
\label{adcoop}
\en 
Note that the vector ${\cal  O}$ is subjected to the
constraints that the set of vectors $n^a$ with $O_{ab}(x) = \{n^1(x),
\, n^2(x),\, n^3(x)\}_{ab}$ is orthonormal. 
At the heart of Laplacian center gauge
fixing, one relaxes these constraints and seeks the $N-1$ largest 
eigenvalues of the super-matrix ${\cal R}$. These tasks can be 
unambiguously performed with present-day algorithmic tools. 
From the corresponding eigenvectors, the adjoint gauge transformations 
$O^{ab}(x)$ are reconstructed at each site with the help 
of Gram-Schmidt orthogonalization. Abelian monopoles and vortices 
appear as defects in the latter step of reconstructing the 
gauge transformation. Technical details of this gauge 
fixing are presented in~\cite{deForcrand:2000pg}. 
Finally, we point out that the Laplacian gauge also seeks to maximize 
the ``mesonic'' gauge condition. However, the re-orthogonalization 
of the eigenvectors implies that the values for the overlap 
$R$ (\ref{overlap}) are significantly smaller than that 
achieved by maximizing $R_{mes}$ (\ref{mmcg}) with the help of 
an iteration over-relaxation procedure.

\subsection{ Identifying vortex matter } 
\label{ident} 

In order to reveal the vortex matter of $Z_N$ gauge theory, we define 
\be 
v(p) \; := \; \prod _{l \in p } Z_l \; , \; \; l=\{x,\mu \} \; , 
\; \; \;  v(p) \in Z_N \; , 
\label{vortex} 
\en 
where $p=(x, \mu < \nu)$ defines an elementary plaquette on the
lattice. 
One says that a vortex of center charge $z$ pierces  the plaquette $p$ if 
\be 
v(p) \; = \; z \; , \hbo 
z \; = \; \exp \Bigl\{ i \frac{ 2 \pi }{N} \; \varphi  \Bigr\} \; ,
\label{charge}
\en 
where $- \frac{N}{2} < \varphi (p) \equiv \varphi _{\mu \nu }(x) 
\le  \frac{N}{2}$ is called the center flux. 
For $SU(N\ge 3)$, one defines the conserved $Z_N$ monopole current by 
\be 
m _\mu (x) \; = \; \frac{1}{N} \; \epsilon _{\mu \nu \alpha \beta } 
\; \Delta _\nu \; \varphi _{\alpha \beta }(x) \; , \hbo 
\Delta _\mu \; m_\mu (x) \; = \; 0 \, ,  
\label{mono}
\en 
where 
$$ 
\Delta _\mu \; \Phi (x) \; :=  \; \Phi (x+\mu) \; - \Phi (x) \; . 
$$
In order to reveal the span of the monopole charge $m_4$, 
we consider an elementary, spatial hypercube 
$c=(x, \alpha < \beta < \gamma)$.  One easily verifies, using the
Abelian nature of the group $Z_N$, that 
\be 
1\; = \; \prod _{p \in c } D\Bigl[ v(p) \Bigr]\; = \; 
\exp \Bigl\{ i \frac{ 2 \pi }{N} \; \epsilon _{ikm} \Delta _i 
\varphi _{km} (x) \; \Bigr\} \; ,
\label{bian}
\en 
where the sum over $i,k,m$ runs from $1\ldots 3$, and 
$ D[ v(p) ] =  v^\dagger (p) $ if the normal vector of the 
plaquette is anti-aligned with the normal vector of the relevant 
surface element of $c$, and  $ D[ v(p) ] =  v(p) $ 
otherwise. Hence, the monopole charge comes in integers, i.e., 
\be 
m_4 \; = \; \frac{1}{N} \; \epsilon _{ikm} \Delta _i \varphi _{km} (x) \; 
= \; k \;,  \hbo k: \, \mathrm{integer} \; . 
\label{cha2}
\en 
Finally, we point out that there are no center monopoles 
in the case of a $SU(2)$ gauge group, i.e., $m_\mu (x) \equiv 0$. 

\vskip 0.3cm 
It is convenient for model building to define the vortex matter 
on the dual lattice, where the link $l$, the plaquette $p$ and 
the cube $c$ is mapped onto 
\be 
l \; \longrightarrow \; c^\ast \; , \hbo 
p \; \longrightarrow \; p^\ast \; , \hbo 
c \; \longrightarrow \; l^\ast \; . 
\label{dlat}
\en 
The vortex field of the dual lattice is defined via the identification 
\be 
\bar{v}(p^\ast ) \; = \; v(p) \; , \hbox to 2cm {\hfill for \hfill } 
p \; \longrightarrow \; p^\ast \; . 
\label{dmap}
\en 
The identity (\ref{bian}) can be transformed into an identity 
for dual fields only: 
\be 
1 \; = \; \prod _{p^\ast \ni l^\ast } \bar{v}(p^\ast ) \, . 
\label{dbian} 
\en 
The latter equation implies that the vortices either from 
closed world sheets on the dual lattice or, for 
$SU(N\ge 3)$ only, multiples of $N$  
vortex world sheets merge at a closed monopole trajectory.

\section{ Dominance of the static quark potential } 
\label{dom} 

Preliminary evidence that the MCG vortex 
matter recovers the string tension of pure $SU(3)$ gauge theory 
was presented in~\cite{Faber:1999sq}. The so-called indirect center 
gauge, i.e., maximal Abelian gauge fixing with a subsequent fixing of 
the center gauge, was investigated  in~\cite{Stack:sy}. 
There it was observed that the string tension obtained from 
Abelian monopole configurations as well as from $Z_3$ vortex ensembles 
is significantly smaller than the string tension of complete $SU(3)$ 
gauge theory. On the other hand, it is interesting to note that the 
$SU(3)$ string tension is recovered to full extent 
in the case of the LCG~\cite{deForcrand:2000pg}. 

\vskip 0.3cm 
Here, we will see that the MCG vortices act like the 
vortices defined in the indirect center gauge: roughly 62\% of 
the full string tension is found. We will therefore find that the 
MCG vortex ensembles genuinely differ from those of the LCG.

\subsection{The case of a $SU(2)$ gauge group revisited } 
\label{sec:su2} 

\begin{figure}
\begin{center} 
\epsfig{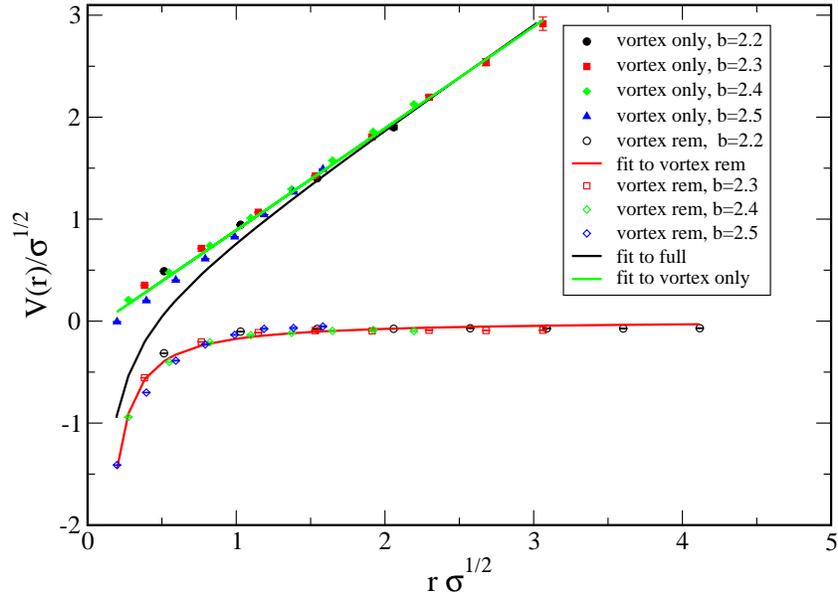}
\end{center}
\caption{ The static quark anti-quark potential for the $SU(2)$ 
gauge group: full ensembles, vortex projected ensembles, and 
ensembles where the vortices have been removed (see (\ref{remov})). 
Lattice size: $24^4$. 
}
\label{fig:3}
\end{figure} 
In order to contrast the findings concerning the 
$SU(3)$ gauge group, which will be presented below, with the 
findings for the case of $SU(2)$, we briefly discuss  
our numerical results for the latter case in this subsection. 

\vskip 0.3cm 
The maximal center gauge (\ref{mmcg}) is implemented by using the 
iteration over-relaxation procedure which is described in detail 
in~\cite{DelDebbio:1998uu}. This procedure defines the gauge. 
The corresponding vortex degrees of freedom are defined by the 
projection (\ref{project}), which becomes in the present case: 
\be 
SU(2) \rightarrow Z_2 : \; \;  U^\Omega _\mu (x) \rightarrow 
Z_\mu (x)  \; = \; \mathrm{sign} \; \tr \;  U^\Omega _\mu (x) \; . 
\label{projectsu2}
\en 
It turns out that this procedure produces vortex configurations 
with sensible properties in the continuum 
limit~\cite{Langfeld:1997jx} and 
with a close relation to the physics of
confinement~\cite{DelDebbio:1998uu,DelDebbio:1996mh,Langfeld:1998cz,Engelhardt:1999fd}.
So far, vortex matter with ``best'' properties in the continuum limit 
seems to obtained with a 'preconditioning' by performing the 
Laplacian center gauge (see subsection \ref{mcg}) and 
subsequent maximal center gauge     
fixing~\cite{Langfeld:2001nz,Faber:2001zs}. This approach 
also alleviates the Gribov problem, but its implementation 
is numerically ``expensive''. 

\vskip 0.3cm 
In order to reveal the relevance of the vortex texture for the 
physics of confinement, one firstly calculates the static 
quark potential from the vortex configurations. 
Secondly, one defines a toy Yang-Mills theory by 
\be 
U^{\prime }_\mu (x) \; = \; Z^\dagger _\mu (x) \; 
U^\Omega _\mu (x) \; , 
\label{remov} 
\en 
where the vortex texture has been removed ``by hand'' from the 
lattice ensembles. It was
found~\cite{DelDebbio:1996mh,DelDebbio:1998uu} that the 
vortex configurations reproduce the linear part of the 
potential to a large extent. In addition, the potential 
evaluated from the modified configurations $\{U^{\prime }_\mu (x) \}$ 
has lost its linear rise and shows a Coulomb type of behavior. 
Both observations are summarized by the term 'center dominance 
of the potential'. 

\vskip 0.3cm 
Figure \ref{fig:3} illustrates these observations using our lattice 
results. Only data which were obtained on a $24^4$ lattice are shown. 
The number of independent configurations employed for the 
calculation of expectation values is listed in table \ref{tab:3}. 
\begin{table}[t]
\begin{center}
\begin{tabular}{c|c|c|c|c} \hline \hline 
$\beta $ & 2.2 & 2.3 & 2.4 & 2.5 \cr \hline 
$\#$ & 75 & 75 & 35 & 55 \cr \hline \hline 
\end{tabular} 
\end{center}
\caption{ Number of independent $SU(2)$ configurations 
used for the calculation of the static potential 
(see figure \ref{fig:3}). } 
\label{tab:3}
\end{table}
The full configurations $\{U_\mu (x) \}$ as well as 
the configurations $\{U^{\prime }_\mu (x) \}$ from which the 
vortices have been removed ``by hand'' are subject to the 
overlap enhancement described in subsection \ref{enhance}. 
The $Z_2$ vortex configurations already possess good 
overlap with the ground state, and no enhancement is used 
in this case. 
The line which fits the ``vortex only'' data in figure \ref{fig:3} 
corresponds to a string tension of 97.7\% of the full string tension. 
we stress that these findings have been obtained with the most 
naive version of the maximal center gauge (described 
in~\cite{DelDebbio:1998uu}). 

\vskip 0.3cm 
Finally, we point out that the quality of dominance is affected 
by the choice of gauge, i.e., the Gribov copy~\cite{Bornyakov:2000ig}, 
and that the Gribov effect is influenced by the lattice 
volume~\cite{Bertle:2000py}.

\subsection{ Vortex-limited Wilson loops: $SU(3)$ gauge group } 
\label{wilson} 

\begin{figure}
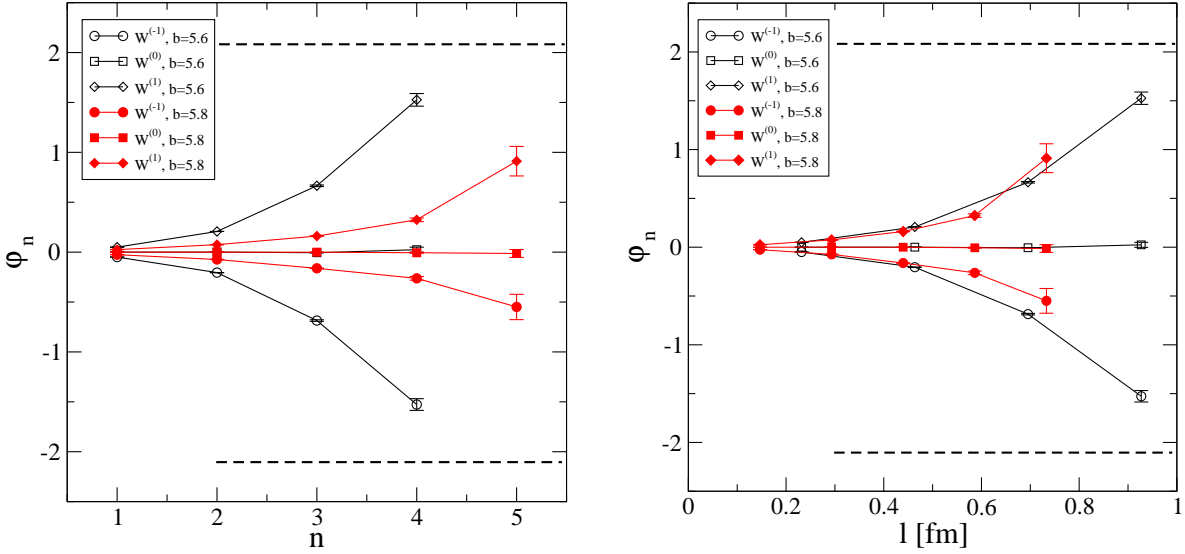

\begin{center} 
\epsfig{figure=phi.eps,width=7.5cm} \hspace{.5cm}
\epsfig{figure=phi_sca.eps,width=7.4cm}
\end{center}
\caption{ The angle $\varphi ^{(m)}$ of the Wilson loop expectation 
values (\ref{angdef}) as a function of the Wilson loop size $n$ 
(left panel) and as a function of the size in physical units (right
panel).  }
\label{fig:4}
\end{figure} 
Let $W_n[U_\mu]$ denote a $n \times n$ planar Wilson
loop calculated within the particular configuration $U_\mu (x)$. 
The same object is 
evaluated with the $Z_3$ configurations obtained from 
center projection (\ref{project}) after direct maximal center 
gauge fixing using the ``mesonic'' gauge condition (\ref{mmcg}). 
The result is called $W_n[Z_\mu]$. Since  $W_n[Z_\mu] \in Z_3$, the 
latter Wilson loop can be characterized by the number 
$m \in \{-1,0,1\}$, i.e., 
\be 
W_n[Z_\mu] \; = \; 
\exp \Bigl\{ i \frac{ 2 \pi }{3} \; m \Bigr\} \; . 
\en 
The expectation value of the Wilson loop, 
\be 
W_n \; = \; \Bigl\langle W_n[U_\mu] \Bigr\rangle \; , 
\en 
is obtained by averaging over the ensembles $\{U_\mu (x) \}$. 
In addition, we can define expectation values 
\be 
W^{(m)}_n \; = \; \Bigl\langle W_n[U_\mu] \Bigr\rangle _{m} \; , 
\en 
where only loops $W_n[U_\mu]$ are taken into account; the 
corresponding quantity $W_n[Z_\mu]$ belongs to the sector 
$m$. Decomposing 
\be 
W^{(m)}_n \; = \; \vert W^{(m)}_n \vert \; \exp \Bigl\{ i \varphi
^{(m)}_n \Bigr\} , 
\label{angdef}
\en 
one expects that for large loops 
\be 
\lim _{n \to \infty } \varphi ^{(m)}_n \; = \; 
\frac{ 2 \pi }{3} \; m \; , 
\label{angul} 
\en 
if center vortices dominate the Wilson loop expectation value. 
The latter relation can be checked by lattice simulations. 
The quantity $\varphi ^{(m)}_n $ is shown for $\beta =5.6$ 
($227$ independent measurements) and for $\beta =5.8$ 
($160$ independent measurements) in figure \ref{fig:4} 
as a function of $n$ (left panel). It seems that the relation 
(\ref{angul}) is indeed satisfied for large $n$. 
If we plot the angle $\varphi ^{(m)}_n $ as a function of the 
physical size of the Wilson loop, i.e., $l = n \, a(\beta )$, 
we observe that the data for  $\beta =5.6$ and for $\beta =5.8$, 
respectively, roughly fall on top of the same curve (right panel). 

\vskip 0.3cm 
Let us interpret these findings from a random vortex model point of
view. Following~\cite{Greensite:2002yn}, we assume that 
center vortex intersection points possess a finite correlation 
length $L_c$. Thus dividing the minimal area $A$ of the planar Wilson 
loop into squares of size ${\cal A} > L_c^2$, the center fluxes 
$\varphi \in \{-1,0,1\} $ through different squares are essentially 
uncorrelated. Let $\widetilde \rho (m,{\cal A})$ denote the 
probability of finding center flux $m$ through the area ${\cal A}$; 
we define the ``mesoscopic'' vortex density by 
\be 
\rho _{mes} \; := \; \widetilde \rho (m,{\cal A}) \; / \; {\cal A} \;
, \hbo A \; = \; n^2 \, a^2(\beta ) \; . 
\label{mesos} 
\en 
Assuming vortex dominance, we might approximate 
\be 
W_n \; = \; \Bigl\langle W_n[U_\mu] \Bigr\rangle 
\; \approx \;  \Bigl\langle W_n[Z_\mu] \Bigr\rangle \; . 
\en 
Using the fact that center fluxes are uncorrelated 
by construction, one obtains 
\be 
W_n \; \approx \;  \Bigl\langle W_n[Z_\mu] \Bigr\rangle \; = \; 
\langle Z \rangle ^{A/{\cal A}} \, , 
\en 
where $ \langle Z \rangle $ the average flux through 
the area ${\cal A}$ is 
\be 
\langle Z  \rangle \; = \; \sum _{m=-1..1} 
 \widetilde \rho (m,{\cal A}) \; \; 
 \exp \Bigl\{ i \frac{ 2 \pi }{3} \; m  \Bigr\} \; . 
\en 
Hence, the string tension in the center flux approximation 
is given by 
\be 
\sigma _{cf} \; = \; - \frac{1}{\cal A} \; \ln \, \Bigl[ 
  \sum _{m=-1..1} \widetilde \rho (m,{\cal A}) \; \;  
 \exp \{ i \frac{ 2 \pi }{3} \; m  \} \Bigr] 
\; = \;  - \frac{1}{\cal A} \; \ln \, \, \Bigl( 1 \; - \; 
3 \;  \widetilde \rho \Bigr) \; , 
\en 
where we have assumed that the center symmetry $m \rightarrow 
-m $ is not spontaneously broken, i.e., 
\be 
\widetilde \rho (1,{\cal A}) \; = \; \widetilde \rho (-1,{\cal A}) 
\; =: \; \widetilde \rho ({\cal A}) \; , 
\hbox to 2cm {\hfill and \hfill } 
\sum _{m=-1}^1 \widetilde \rho (m,{\cal A}) \; = \; 1 \; . 
\en 
A particular case is obtained by considering that the vortices 
which are defined at the level of the elementary plaquette are 
uncorrelated (naive random vortex model). 
In this case, one finds 
\be 
 \widetilde \rho ({\cal A}=a^2) \; = \; \frac{\rho }{2} \; a^2 
\; \ll \; 1 \, , \hbox to 3cm {\hfill and thus \hfill } 
\sigma _{cf} \; \approx \; \frac{3}{2} \; \rho \; . 
\label{naive}
\en 
Thereby,  $\rho $ is the ``microscopic'' vortex density, i.e., 
$\rho a^2$ is the probability of finding a non-trivial center 
flux through a given plaquette (no matter whether $m=-1$ or $m=1$). 

\vskip 0.3cm 
Using the numerical data above, it is possible to estimate 
the center flux correlation length. Let us define the 
``half-width'' $L_{1/2}$ by the length of the Wilson loop at which 
\be 
\varphi ^{(1)} (l=L_{1/2}) \; = \; \frac{\pi }{3} \, . 
\en 
The findings (see figure \ref{fig:4}, right panel) suggest that 
\be 
L_c \; \ge \; L_{1/2} \; \approx \; 0.8 \; \mathrm{fm} \, , 
\label{lc}
\en 
where we have used a string tension of $\sigma = (440 \;
\mathrm{MeV})^2 $ as a reference scale. 
Finally, let us check whether the naive random vortex model 
of uncorrelated vortex plaquettes is realistic. For this to be the 
case, the relation  
\be 
\frac{ \rho }{2} \; {\cal A } \; \approx \; \frac{ \rho }{2} \; 
L^2_{1/2}  \; \approx \; \frac{ \sigma _{cf}  }{3} \; L^2_{1/2} 
 \approx \;  \frac{ \sigma  }{3} \; L^2_{1/2} \; \ll \; 1 
\en 
must hold. However, one finds, using (\ref{lc}), 
\be 
\frac{ \sigma  }{3} \; L^2_{1/2} \; \approx \; 1.03 \; , 
\en 
implying that the naive random vortex model seems not 
always to be justified.

\subsection{ The ``mesonic'' center gauge for $SU(3)$ } 
\label{potmes} 

\begin{figure}[t]
\begin{center} 
\epsfig{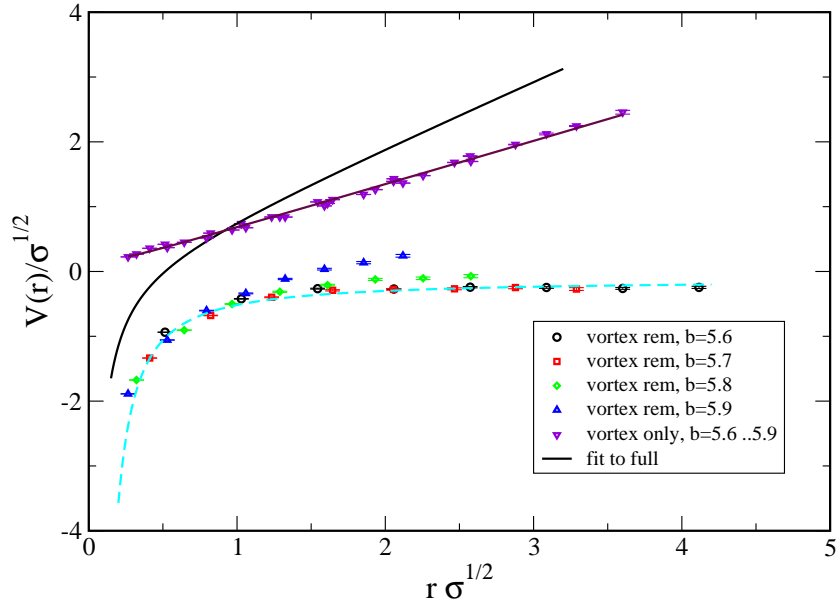}
\end{center}
\caption{ The static quark anti-quark potential for the $SU(3)$ 
gauge group: full ensembles, vortex projected ensembles, and 
ensembles where the vortices have been removed (see (\ref{remov})). 
Lattice size: $16^4$. 
}
\label{fig:5}
\end{figure} 
In a first step, the ``mesonic'' gauge condition (\ref{mmcg}) 
is installed with the help of the iteration over-relaxation 
algorithm described in detail in~\cite{Faber:1999sq,Montero:1999by}. 
The $Z_3$ link elements $Z_\mu(x)$ are defined by center projection 
(\ref{project}). As in the case of $SU(2)$, we will compare the 
static quark potential obtained from full link configurations 
(see section \ref{pot}) with the one calculated with link 
ensembles $\{Z_\mu (x)\}$. In addition, the toy model 
is defined by configurations $\{U ^\prime _\mu (x)\}$ (\ref{remov}) 
from which the vortices have been removed ``by hand''. From the
results of the previous subsection, we expect that the 
string tension is lost in the latter case. 
Our numerical findings using $100$ independent measurements are 
summarized in figure \ref{fig:5}. 

\vskip 0.3cm 
We find that the potential calculated from vortex configurations 
scales towards the continuum limit, i.e., the data 
obtained from different $\beta $ values fall on top of the same 
curve if the $V(r)$ and $r$ are expressed in physical units. 
In addition, one observes ``precocious'' linearity: the potential 
is linear even at small distances as is the case for a 
$SU(2)$ gauge group. In contrast to the case of an $SU(2)$ gauge 
group, the center projected string tension is only 62\% 
of the full string tension. The value of string tension (in lattice 
units) after center projection is in agreement with the finding 
in~\cite{Faber:1999sq} for a $12^4$ lattice and $\beta =5.6$. 
In the latter article, however, the quoted value of the full 
string tension is underestimated. Using reliable values, the 
ratio of projected and full string tension is in agreement 
with the findings reported here. 

\vskip 0.3cm 
On the other hand, removing the vortices 
(see (\ref{remov})) produces configurations which are compatible 
with a vanishing  string tension. There is a subtlety for 
obtaining this result: the lattice volume must be large 
enough\footnote{I thank M.~Faber for this remark.}.
It appears that the lattice size of $16^4$ seems to be too small 
for $\beta $ as big as $6.0$. 

\begin{figure}
\begin{center} 
\epsfig{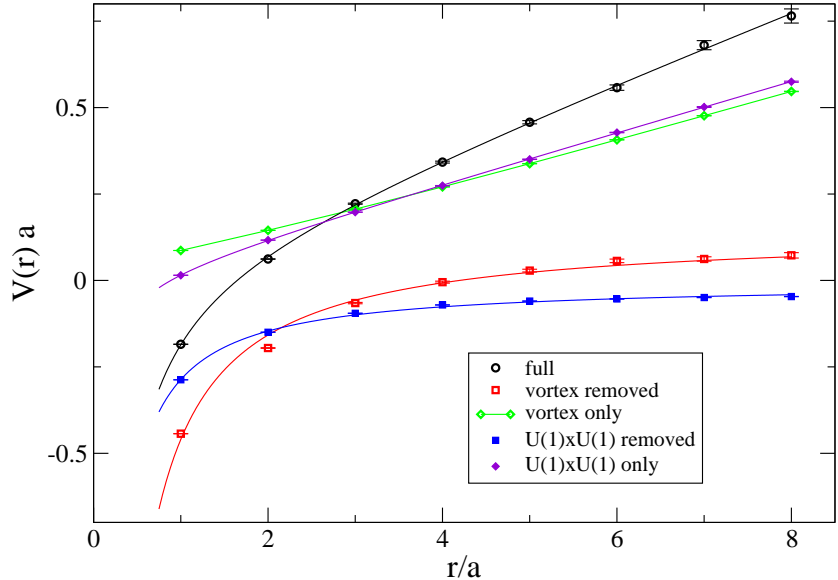}
\end{center}
\caption{ The static quark anti-quark potential for the $SU(3)$ 
gauge group: full ensembles, vortex projected ensembles, and 
ensembles where the vortices have been removed (see (\ref{remov})). 
Same with $U(1)\times U(1)$ projection. 
Lattice size: $16^4$, $\beta =5.8$. 
}
\label{fig:6}
\end{figure} 
\vskip 0.3cm 
Since a removal of the vortices results in a loss of the string
tension, even if the vortices only amount to 62\% of the full one, 
the question arises as to whether additional degrees of freedom which
reside in the $U(1) \times U(1)$ Abelian subgroup are responsible 
for the 38\% string tension completing the vortex contribution. 
Candidates for such degrees of freedom are color magnetic monopoles. 
To answer this question, we implemented the ``mesonic'' gauge 
condition (\ref{mmcg}) and subsequently projected the 
gauged configurations onto Abelian ones: 
\be 
U^\Omega _\mu (x) \longrightarrow V_\mu (x) \; , \hbo 
 V_\mu (x) \in U(1) \times U(1) \; . 
\label{ablian} 
\en 
For these purposes, the off-diagonal elements of $U^\Omega _\mu (x) $ 
were dropped, i.e., 
\be 
U^\Omega _\mu (x) \longrightarrow \bar{U}^\Omega _\mu (x) 
:= \mathrm{diag}\Bigl( U^\Omega _\mu (x) \Bigr) \; , 
\en 
and $V_\mu (x)$ is given by the $SU(3)$ element which is ``closest'' 
to $ \bar{U}^\Omega _\mu (x) $ (see discussion in subsection 
\ref{enhance}). One verifies that indeed $ V_\mu (x) \in U(1) \times
U(1) $. In addition, we investigated ensembles $\{U ^{ABR} _\mu (x)\}$ 
which are complementary to the $U(1) \times U(1)$ configurations: 
\be 
U ^{ABR} _\mu (x) \; = \; V^\dagger _\mu (x)  \; U^\Omega _\mu (x) 
\; . 
\label{abrem} 
\en 
Our numerical findings for a $16^4$ lattice at $\beta =5.8$ are 
shown in figure \ref{fig:6}. 
we find that the string tension calculated from $U(1) \times U(1)$ 
configurations is marginally larger than the string tension from 
vortex projected configurations. 
As expected, configurations $U ^{ABR} _\mu (x)$ 
from which the Abelian subgroup was removed do not support confinement. 

\begin{figure}
\begin{center} 
\epsfig{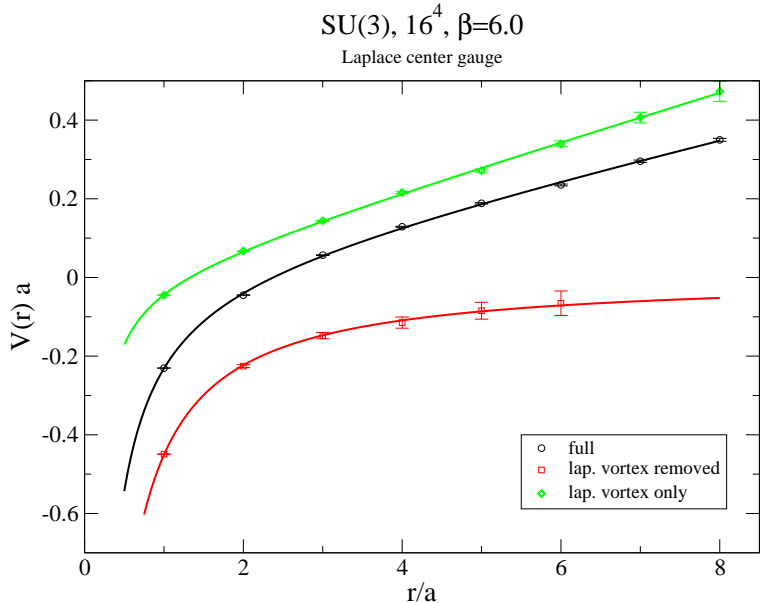}
\end{center}
\caption{ The static quark anti-quark potentials for the $SU(3)$ 
gauge group using the Laplacian center gauge: full ensembles, 
vortex projected ensembles, and 
ensembles where the vortices have been removed (see (\ref{remov})). 
Lattice size: $16^4$, $\beta =6.0$. 
}
\label{fig:7}
\end{figure} 
\vskip 0.3cm 
These results are compared with those in~\cite{Stack:sy}: 
there, versions of the so-called Maximal Abelian Gauge were 
investigated. These gauges are most suitable for a projection of 
configurations $U^\Omega _\mu (x) $ onto the Abelian subgroup 
$U(1) \times U(1)$. Also in these cases, the string tension 
extracted from $U(1) \times U(1)$ configurations is substantially 
smaller than the full string tension.

\subsection{ Laplacian center gauge } 
\label{lap_pot} 

The previous subsection showed that the ``mesonic gauge'' 
(\ref{mmcg}) produces vortex matter which recovers only 
62\% of the full string tension. The question is whether 
$Z_3$ matter is able to give the full result for the string 
tension at all. The answer was already given in~\cite{deForcrand:2000pg}: 
vortex matter which is defined by the Laplacian gauge condition 
(see subsection \ref{mcg}) reproduces the linear rise of 
the static quark potential in the continuum limit. 
Here, we briefly report our findings. We investigated the 
somewhat extreme case of a small physical volume, i.e., 
$16^4$ lattice, $\beta =6.0$. The last subsection showed that 
for this size the removal of the vortices defined by the ``mesonic 
gauge'' hardly makes the string tension vanish. The result 
of $50$ independent measurements is shown in figure \ref{fig:7}. 
The potentials of this figure were fitted by the function 
\be
V(r) \; a^2 \; = \; \sigma \; a^2 \; n \; \; - \; \; \frac{\alpha }{n} 
\; , \hbo r = n a \, . 
\en  
In either case we find 
\bea 
\mathrm{ vortex \; only } \; &:& \; \sigma a^2 \; = \; 0.061 \; 
\hbo \alpha \; = \; 0.096 \; , \nonumber \\ 
\mathrm{ vortex \; removed } \; &:& \; \sigma a^2 \; = \; 0.0\phantom{61}\; 
\hbo \alpha \; = \; 0.43 \; , \nonumber \\ 
\mathrm{ full } \; &:& \; \sigma a^2 \; = \; 0.051 \; 
\hbo \alpha \; = \; 0.25 \; . \nonumber 
\ena 
A small Coulomb part survives the projection onto vortices. 
One also observes that the potential obtained from configurations 
from which the Laplacian vortices have been removed 
is perfectly fitted by a Coulomb law. The string tension from 
vortex configurations is a bit higher than the full string tension. 
This is probably due to the small physical size of the lattice: 
Coulomb contributions are dominant and are represented by the 
vortex matter as string tension to some extent. 

\vskip 0.3cm 
It turns out~\cite{Langfeld:2001nz,Faber:2001zs}
in the case of the $SU(2)$ gauge group 
that preconditioning with the Laplace center gauge 
and subsequent implementation of the ``mesonic'' gauge 
by iteration over-relaxation strongly reduces the influence of the 
Gribov copies. Center projection of these ensembles yields 
high-quality vortex matter the properties of which nicely 
extrapolates to the continuum limit. 
We have checked that repeating this approach for the $SU(3)$ case 
produces vortex matter where the corresponding string tension 
again only reaches 62\% of the full string tension.

\section{Properties of $SU(3)$ vortex matter } 
\label{properties} 

\begin{figure}
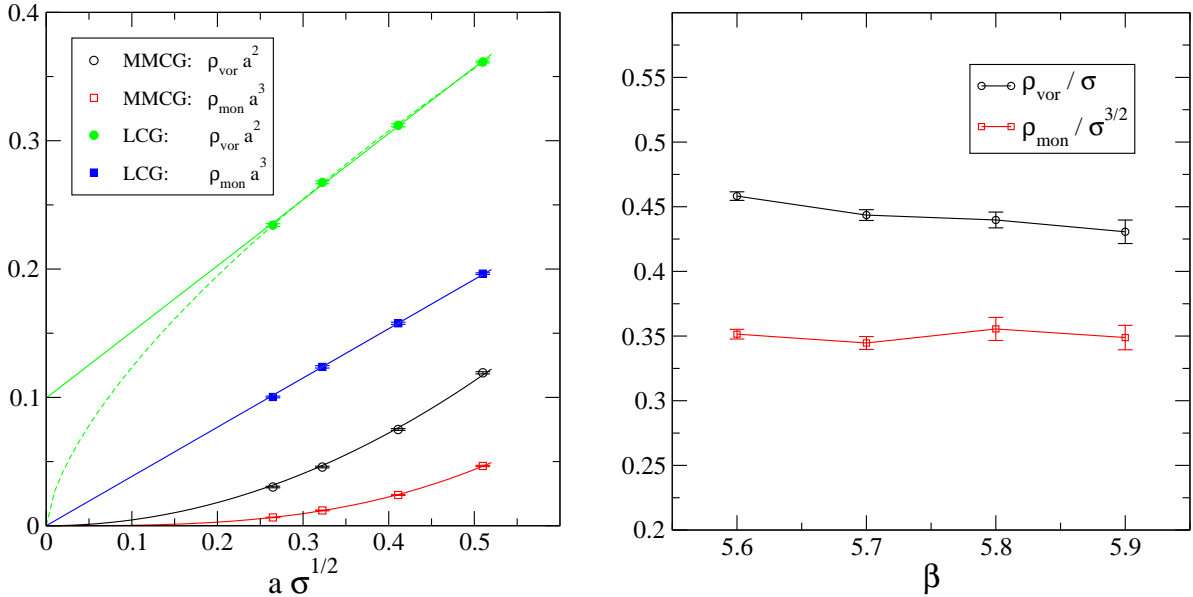

\begin{center} 
\epsfig{figure=scaling.eps,width=7.4cm} \hspace{.5cm}
\epsfig{figure=vdens.eps,width=7.6cm}
\end{center}
\caption{ The continuum limit of the vortex- and the $Z_3$ monopole 
density for the ``mesonic'' gauge and the Laplacian gauge (left
panel). The vortex- and the $Z_3$ monopole density in physical units 
for the ``mesonic gauge'' (right panel). 
}
\label{fig:8}
\end{figure} 
In this section, we will investigate which definition of the 
vortex matter (MCG or LCG) admits an interpretation of the vortices 
as physical degrees of freedom. To this aim, the continuum limit 
$a \rightarrow 0$ of the planar vortex density $\rho _\mathrm{vor}$ 
(i.e., the ``microscopic'' vortex density of 
subsection \ref{wilson}) and the density $\rho _\mathrm{mon}$ of $Z_3$ 
monopoles are investigated.

\vskip 0.3cm 
The vortex density $\rho _\mathrm{vor}$ can be extracted from 
the $Z_3$ ensembles $\{Z_\mu (x)\}$ as follows: we say that 
a $Z_3$ plaquette, 
\be 
v(p) \; := \; \prod _{l \in p } Z_l \; , \; \; l=\{x,\mu \} \; , 
\; \; \;  v(p) \in Z_3 \; , 
\label{vortex3} 
\en 
carries a non-trivial center flux if 
\be 
v(p) \; = \; z \; , \hbo 
z \; = \; \exp \Bigl\{ i \frac{ 2 \pi }{3} \; \varphi  \Bigr\} 
\label{charge3}
\en 
and $\varphi = -1 $ or $\varphi = 1 $. 
If ${\cal P}_\beta $ denote the probability that a particular 
plaquette of the lattice carries a non-trivial center charge, 
the vortex density is defined by 
\be 
\rho _\mathrm{vor} \; a^2 (\beta ) \; = \; {\cal P}_\beta \; . 
\label{vorden} 
\en 
As outlined in subsection \ref{ident}, the $Z_3$ monopole world lines 
are associated with dual links whose corresponding monopole current 
$m_\mu (x)$ (\ref{mono}) is non-zero. Alternatively, 3d hypercubes 
from which non-trivial center flux emerges are said to contain 
a center monopole. Given that ${\cal Q}_\beta $ is the probability 
that a particular hypercube contains a center monopole, the monopole 
density is obtained from 
\be 
\rho _\mathrm{mon} \; a^3 (\beta ) \; = \; {\cal Q}_\beta \; . 
\label{monden} 
\en 
Both quantities characterize the vortex matter. In order to 
interpret the vortices as sensible degrees of freedom in the 
continuum limit $a \rightarrow 0$, the quantities must obey 
\be 
\lim _{\beta \to \infty } \rho _\mathrm{vor} \; = \; \mathrm{constant}
\; , \hbo 
\lim _{\beta \to \infty } \rho _\mathrm{mon} \; = \; \mathrm{constant}
\; . 
\en 
Our numerical findings are summarized in figure \ref{fig:8}. 
Figure \ref{fig:8}, left panel shows the densities $\rho
_\mathrm{vor}$ and  $\rho _\mathrm{mon}$ in units of the lattice 
spacing as a function of $a \; \sqrt{\sigma }$. Simulations were 
performed for $\beta = 5.6$, $5.7$, $5.8$, $5.9$. The corresponding 
size of the lattice spacing can be found in table \ref{tab:2}. 

\vskip 0.3cm 
Let us firstly focus on vortex matter obtained after implementing 
the ``mesonic'' gauge condition (\ref{mmcg}). There, the data 
are perfectly fitted by 
\bea 
 \rho _\mathrm{vor} \; a^2 (\beta ) &\approx&  0.45 \; \Bigl( a \; \sqrt{\sigma
 } \Bigr) ^2 \; , \\ 
 \rho _\mathrm{mon} \; a^3 (\beta ) &\approx &  0.35 \; \Bigl( a \; \sqrt{\sigma
 } \Bigr) ^3 \; . 
\ena 
These findings suggest that the planar areal vortex density 
as well as the $Z_3$ monopole density properly extrapolate to 
the continuum limit (see figure \ref{fig:8}, right panel). 
The same result for the vortex density was found for the case 
of a $SU(2)$ gauge group~\cite{Langfeld:1997jx,DelDebbio:1998uu}. 
In a naive random vortex model, one expects that the 
string tension is given by (see (\ref{naive}))
\be 
\sigma _{cf} \; \approx \; \frac{3}{2} \; \rho _\mathrm{vor} \; 
\approx \; 0.67 \; \sigma \; . 
\en 
We point out that the 
naive string tension $\sigma _{cf}$ roughly agrees with 
the string tension measured from center projected configurations 
(see subsection \ref{potmes}), i.e., 
$$ 
\sigma _{Z3} \; \approx \; 0.62 \; \sigma \; . 
$$ 
This indicates that the intersection points of the ``mesonic'' center 
vortex clusters do not contain significant long-range correlations. 
Using $\sqrt{\sigma }= 440 \,$MeV as a reference scale, one finds 
for the SU(3) case
\be 
 \rho _\mathrm{vor} \; \approx \; 2.2 \; \frac{1}{\mathrm{fm}^2} \; , 
\hbo 
 \rho _\mathrm{mon} \; \approx \; 3.7 \; \frac{1}{\mathrm{fm}^3} \; , 
\hbo 
\frac{  \rho _\mathrm{mon} }{  \rho ^{3/2} _\mathrm{vor} }
\; \approx \; 1.16 \; .  
\en 
The latter quantity might be of interest for the construction of 
$Z_3$ random vortex models.

\vskip 0.3cm 
As already noticed for a $SU(2)$ gauge group~\cite{Langfeld:2001nz}, 
the situation drastically changes for the Laplacian center gauge. 
The vortex and monopole densities (times the canonical powers 
of the lattice spacing) scale linearly with the lattice spacing. 
Two fits represent the areal vortex density almost equally well: 
\be 
 \rho _\mathrm{vor} \; a^2 (\beta )  \; \approx \; 0.1 \; + \; 0.52 \; 
a \; \sqrt{\sigma } \; ,  \hbo 
 \rho _\mathrm{vor} \; a^2 (\beta )  \; \approx \; 0.56 \; 
\Bigl( a \; \sqrt{\sigma } \Bigr) ^{0.66} \; , 
\en 
where the latter fit function is slightly preferable.
In addition, the monopole density is well represented by the linear 
function
\be 
 \rho _\mathrm{mon} \; a^3 (\beta ) \; \approx \;  0.38 \;  a \; \sqrt{\sigma
 } \; . 
\en 
Both quantities, i.e., $ \rho _\mathrm{vor}$ and  $\rho
_\mathrm{mon}$,  
diverge in the continuum limit $a \rightarrow 0$. 
However, one finds that the ratio 
\be 
\frac{  \rho _\mathrm{mon} }{  \rho ^{3/2} _\mathrm{vor} }
\; \approx \; 0.9 \hbox to 5cm {\hfill (LCG) }
\en 
is roughly independent of the lattice spacing $a$, as is the case 
for the ``mesonic'' gauge.  

\vskip 0.3cm 
\begin{figure}
\begin{center} 
\epsfig{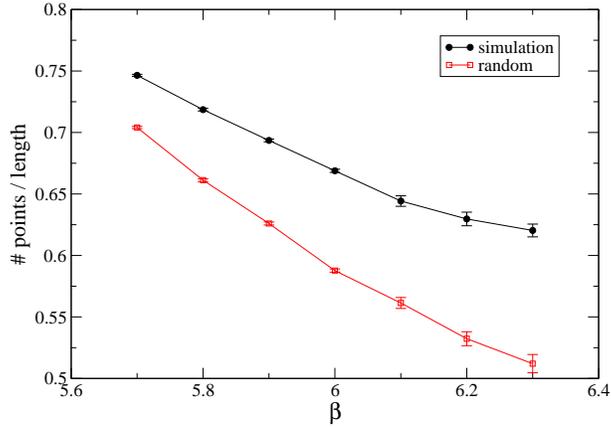}
\end{center}
\caption{ (line) density of monopoles on the 
shortest connected line joining all monopole sites: 
lattice simulation (full symbols) and after randomly re-distributing 
the monopoles (open symbols). 
}
\label{fig:9}
\end{figure} 
It is interesting that the $Z_3$ monopole density (of the Laplacian
gauge) diverges in a somewhat controlled 
way: a situation where the $Z_3$ monopoles lie dense on 2d
hypersurfaces of the 4d space-time would correspond to the observed 
scaling with the lattice spacing. In order to get a rough idea of how the 
the monopoles are organized within space, a closed loop which joins all 
monopole sites is calculated. This loop is obtained by a simulated annealing
procedure which minimizes the length of the loop. Thereby, the 
Euclidean norm (where in addition the toroidal topology is taken into 
account) is used as a measure for lengths. From a
mathematical point of view, finding the global minimum is 
a ``traveling salesman'' problem in three dimensions, and is beyond 
the reach of numerical calculation. However, the ``simulated
annealing'' algorithm generically generates paths the length 
of which is within a few percents of the minimal length. 
This suffices for our purposes here. Given a finite set of points, it is
difficult to tell whether the points are falling on top of a
``smooth'' curve. In order to gain first insight, we have calculated 
a connecting loop for a given set of $Z_3$ monopoles with the help 
of simulated annealing. Dividing the number of monopoles by the length 
of this loop gives an effective line density. Finally, this
quantity is averaged over several lattice configurations. 
In order get a clue about the significance of this average line
density, we have randomly re-distributed the monopoles of a particular 
configuration and we have re-calculated the effective line density. 
If the monopoles produced by the lattice simulation tend to fall on 
top of a smooth line, the average line density must be significantly 
larger than in the case of the random distribution of the same 
amount of monopoles. This is indeed the case, as is shown by 
figure \ref{fig:9}.

\section{Conclusions}
\label{Conclu}

MCG vortices of the gauge group $SU(2)$ have been realized 
as sensible degrees of freedom in the continuum
limit~\cite{Langfeld:1997jx}, and they are closely related to
confinement~\cite{Greensite:2003bk}. 
In the present paper, the MCG vortex matter has been investigated 
for the important case of  $SU(3)$, using large-scale numerical 
simulations. Focal points were the questions: 
To which extent are the $SU(3)$ vortices relevant for confinement? 
Are the $SU(3)$ vortices meaningful in the continuum limit?

\vskip 0.3cm 
In a first step, we verified that the phase of large Wilson loops 
is represented by the center flux going through the Wilson loop. 
we confirmed the conjecture in~\cite{Greensite:2002yn} that the $Z_3$ fluxes
going through a planar area are strongly correlated at length scales 
smaller than the hadronic one. As a byproduct, the flux correlation 
length (see (\ref{lc}) for a proper definition) 
$$
L_{1/2} \; \approx \; 0.8 \; \mathrm{fm} 
$$
was seen to be in rough agreement with scaling. 

\vskip 0.3cm 
In a second step, the static quark potential produced by the 
MCG vortices was addressed. In contrast to the case of $SU(2)$, 
the string tension from the vortex projected ensembles 
turned out to be 62\% of the full string tension. 
This finding is rather independent of the lattice size and the 
value of the lattice spacing. On the other hand, removing the $Z_3$
vortex  degrees of freedom ``by hand'' from the full lattice configurations 
(see (\ref{remov})), always results in a vanishing string tension 
(if the physical lattice volume is large enough). This implies that 
there is still a certain relation between the MCG vortices and 
confinement. 

\vskip 0.3cm 
On the other hand, it is
known~\cite{Alexandrou:1999iy,deForcrand:2000pg} that vortices 
which are defined from the Laplacian center gauge (LCG) 
reproduce the string tension to full extent. 
Here, we checked that ``preconditioning'' the lattice 
configurations with LCG and subsequent MCG fixing does not 
produce vortex matter which yields significantly more than 
62\% string tension. 

\vskip 0.3cm 
The question arose as to which definition (MCG or LCG) of the vortices
produces vortex structures which are sensible in the continuum 
limit. Here,  the planar vortex density $\rho_\mathrm{vor}$ (the
density of points where the vortices intersect a 2D planar
hypersurface) as well as the (volume) density $\rho _\mathrm{mon}$ 
of $Z_3$ monopoles were studied. we found that in the case 
of the MCG vortices both quantities properly extrapolate 
to the continuum: 
$$ 
 \frac{ \rho _\mathrm{vor} }{ \sigma } \; \approx \;   0.45 \; , \hbo 
 \frac{ \rho _\mathrm{mon} }{ \sigma ^{3/2} } \; \approx \;    0.35 \;
 , \hbo 
\frac{  \rho _\mathrm{mon} }{  \rho ^{3/2} _\mathrm{vor} }
\; \approx \; 1.16 \; .  
$$ 
In contrast, both quantities diverge in the continuum limit 
for the case of LCG vortices. Surprisingly, the LCG vortex matter 
satisfies simple scaling laws: 
$$ 
 \rho _\mathrm{mon} \; a^3 (\beta ) \; \approx \;  0.38 \;  a \; 
\sqrt{\sigma } \; , \hbo 
\frac{  \rho _\mathrm{mon} }{  \rho ^{3/2} _\mathrm{vor} }
\; \approx \; 0.9 \hbox to 2cm {\hfill (LCG) } \; . 
$$ 
The investigations of sets of $Z_3$ monopoles residing within 
the spatial hypercube indicated that the LCG monopoles tend to fall 
on top of a ``smooth'' 1d curve which is embedded in this hypercube. 

\vskip 0.3cm 
In summary, only the maximal center gauge allows for a direct 
interpretation of the vortices in the continuum limit of pure $SU(3)$ 
gauge theory. There is no string tension without MCG vortices, but MCG 
vortices only support 62\% of the full value.

\section*{Acknowledgment}
I am grateful to M.~Faber, J.~Greensite, M.~E.~Ilgenfritz, 
D.~J.~Kusterer, S.~Olejnik, M.~Quandt and H.~Reinhardt for helpful 
discussions. I thank F.~Pederiva and B.~Rossi for the help in 
using the FEP computer cluster at the ECT, Trento, where parts 
of the numerical calculations were performed.

\end{document}